\documentclass[aps,prl,twocolumn]{revtex4}
\usepackage{amssymb}
\usepackage{graphicx}
\usepackage{amsmath}
\usepackage{subfigure}
\usepackage{epsfig}
\usepackage{float}
\usepackage{color}
\usepackage{lscape}
\usepackage{pdflscape}
\usepackage[figuresright]{rotating}
\usepackage{changes}

\begin{document}

\title{A Hunt for Highly Charged Ions as Ultra-stable Optical Clock Candidates: Art of the energy level-crossing approach}

\author{$^a$Yan-mei Yu}
\email{ymyu@aphy.iphy.ac.cn}

\author{$^b$B. K. Sahoo}
\email{bijaya@prl.res.in}

\affiliation{$^a$Beijing National Laboratory for Condensed Matter Physics, Institute of Physics, Chinese Academy of Sciences, Beijing 100190,China\\
$^b$Atomic, Molecular and Optical Physics Division, Physical Research Laboratory, Navrangpura, Ahmedabad 380009, India }

\date{\today}

\begin{abstract}
We examine energy level-crossings of fine-structure (FS) levels in the heavier highly charged ions (HCIs) with $d^6$ and $d^8$ configurations. From the
analysis, we find that some of these HCIs are tailor-made for atomic clocks with quality factors ranging from $10^{16}$ to $10^{18}$ and fractional
uncertainties below $10^{-19}$ level. Many of them are also realized to be highly sensitive to the observation of temporal variation of FS
constant ($\alpha$) and testing violation of local Lorentz symmetry invariance (LLI). To probe variations of $\alpha$ and LLI, we have determined
their corresponding sensitivity coefficients in the investigated HCIs. Similarly, we have estimated orders of magnitudes of the Zeeman, Stark, black-body
radiation and electric quadrupole shifts of the clock transitions of the considered HCIs in order to demonstrate them as the potential candidates for
atomic clocks.
\end{abstract}

\maketitle

Highly charged ion (HCI) as atomic clock using the magnetic dipole (M1) transition between the fine-structure (FS) splitting of the ground state in Ar$^{13+}$ has been realized in laboratory \cite{King-nature-2022}. The motivation to use Ar$^{13+}$ as an ultra-precise optical clock was driven from the theoretical studies carried out earlier \cite{Yudin-PRA-2014,Yu-PRA-2019}. Generally, the HCIs have enhanced sensitivity coefficients to variation of many fundamental
physical constants and they show least response to external perturbations. These unique characteristics of HCIs make them potential candidates, when
considered as optical clocks, as the finest sensors to test many fundamental postulates of modern quantum mechanics \cite{Safronova2018,Kozlov2018,Yu2023}.
In view of these testings, it is essential to use heavier HCIs than Ar$^{13+}$ ion for optical clocks. Also, improvement in the uncertainties of
Ar$^{13+}$ clock could be a challenge.

\begin{table}[t]
\caption{Comparison of experimental (Exp.) and calculated values of excitation energies (in cm$^{-1}$) of the fine-structure levels in
the Cr-like, Fe-like, Mo-like, W-like and Ru-like HCIs with $d^6$ and $d^8$ configurations. Calculations are carried out at
the CI+MBPT, CCSD, CISD, and CISDT method approximations but results from the CISDT method are considered as our final theoretical
results for which differences from the Exp. values \cite{NIST,Raassen-PC-1986} are given as Diff. in percentage. \label{tab:EEcomp}}
	{\setlength{\tabcolsep}{2pt}
\begin{tabular}{cccccccc }\hline\hline
Level	&	Exp. 	& CI$+$MBPT	&	CCSD	&	CISD	&	CISDT	&	Diff.(\%)	\\ \hline
\multicolumn{6}{l}{Cr-like Zn$^{6+}$ with $3d^6~^5D_4$ ground state \cite{NIST} }														\\
$^5D_3$	&	1567 	&	1595 	&		&	1559 	&	1572 	&	0.3 	\\
$^5D_2$	&	2579 	,&	2633 	&		&	2586 	&	2584 	&	0.2 	\\
$^5D_1$	&	3230 	&	3294 	&		&	3239 	&	3226 	&	$-0.1$ 	\\
$^5D_0$	&	3542 	&	3621 	&		&	3557 	&	3537 	&	$-0.1$ 	\\ \hline
\multicolumn{6}{l}{Fe-like Rb$^{11+}$ with $3d^8~^3F_4$ ground state  \cite{NIST}}													\\
$^3F_3$	&	10980	&	10980 	&	10772 	&	11276 	&	11285 	&	2.8 	\\
$^3F_2$	&	15610	&	15796 	&	15468 	&	16278 	&	15978 	&	2.4 	\\
$^1D_2$	&	34020	&	35410 	&	34571 	&	37159 	&	35284 	&	3.7 	\\
$^3P_1$	&	46580	&	48774 	&	46748 	&	50590 	&	48642 	&	4.4 	\\
$^3P_0$	&	47220	&	49417 	&	47318 	&	51362 	&	49130 	&	4.0 	\\
$^3P_2$	&	48070	&	49654 	&	48731 	&	52037 	&	50034 	&	4.1 	\\ \hline
\multicolumn{6}{l}{Mo-like Pd$^{4+}$ with $4d^6~^5D_4$ ground state \cite{Raassen-PC-1986}}														\\
$^5D_3$	&	2103.8 	&	2167 	&		&	2106 	&	2136 	&	1.6 	\\
$^5D_2$	&	3175.3 	&	3330 	&		&	3259 	&	3239 	&	2.0 	\\
$^5D_1$	&	3950.0 	&	4141 	&		&	4053 	&	4027 	&	1.9 	\\
$^5D_0$	&	4306.4 	&	4526 	&		&	4429 	&	4391 	&	2.0 	\\ \hline
\multicolumn{6}{l}{W-like Pt$^{4+}$ with $5d^6~^5D_4$ ground state \cite{NIST}}													\\
$^5D_2$	&	6026.34 	&	6712 	&		&	6720 	&	6337 	&	5.2 	\\
$^5D_3$	&	7612.63 	&	7576 	&		&	7210 	&	7412 	&	$-2.6$ 	\\
$^5D_0$	&	10817.6 	&	11378 	&		&	11216 	&	10893 	&	0.7 	\\
$^5D_1$	&	10826.8 	&	11085 	&		&	10797 	&	10756 	&	$-0.7$ 	\\ \hline
\multicolumn{6}{l}{Ru-like Xe$^{10+}$ with $4d^8~^3F_4$ ground state \cite{NIST}}													\\
$^3F_2$	&	13140.	&	13488 	&	13453 	&	14250 	&	13322 	&	1.4 	\\
$^3F_3$	&	15205.	&	14986 	&	15472 	&	15490 	&	15114 	&	$-0.6$ 	\\
$^3P_2$	&	26670.	&	26934 	&	27172 	&	27884 	&	26812 	&	0.5 	\\
$^3P_0$	&	32210.	&	33192 	&	32924 	&	34715 	&	32637 	&	1.3 	\\
$^3P_1$	&	34610.	&	35396 	&	35225 	&	36462 	&	34879 	&	0.8 	\\ \hline
\hline	
\end{tabular}}
\end{table}

HCIs with large $Q$ are particularly useful to overcome low signal-to-noise ratio and to improve the stability limit. For example, though Ar$^{13+}$ has
a relatively simple electronic structure, its milli-Hz clock linewidth impedes the clock stability. This could result in longer averaging time up to days
to achieve the intended high accuracy. As outlined in Ref. \cite{Yu2023}, one of the most prevailing approaches to find out suitable atomic clock
candidates is to analyze energy level-crossings (ELCs) in the isoelectronic HCIs. Historically, study of ELCs in atomic systems are interesting to fathom
energy level shifting in the presence of external electromagnetic fields \cite{Robiscoe-PR-1965,Levine-RPL-1969}. In case of HCIs, this phenomena refers
to the reordering of energy level positioning with the degree of ionization \cite{Berengut-RPL-2010,Berengut-PRA-2012}. By analyzing ELCs, a number of
HCIs have been proposed as potential candidates for atomic clocks like Ir$^{17+}$ \cite{Berengut-PRL-2011}, Pr$^{9+}$ \cite{Bekker-NC-2019}, Cf$^{16+}$
and Cf$^{17+}$ \cite{Berengut-PRL-2012, Porsev-PRA-2020}, Nd$^{9+}$ \cite{Yu-atom-2022}, etc.. Near ELCs, frequencies of atomic transitions can be within
the optical range which can be used as frequency standards. Accurate prediction of energy level structures and their spectroscopic properties in these HCIs, especially
those are having multivalence electrons in the outer $f$-shell, are extremely difficult using available many-body methods.

In this Letter, we demonstrate a unique class of heavier HCIs as potential candidates for optical clocks having larger $Q$ values with projected
fractional uncertainties below 10$^{-19}$ level. These HCIs have $(n=4,5)d^{6,8}$ electronic configurations giving rise to multiple FS splitting. By
analyzing ELC of these FS levels (FS-ELC) with the increasing values of atomic number ($Z$) and ion charge ($Z_{\rm ion}$) along an
isoelectronic sequence, we shortlist a few selective HCIs possessing at least two optical-accessible clock transitions. Simultaneous interrogation
of two clock transitions in a given atomic system may be useful to minimize systematic uncertainties in clock experiments like in Yb$^{+}$
\cite{Lange-PRL-2021} to probe variation of FS constant ($\alpha$). Moreover, the clock transitions in these HCIs show negative differential values for
dipole polarizabilities ($\alpha^{E1}$) suggesting that these ions can also be advantageous for undertaking in the multi-ion clock scheme
\cite{Arnold-PRA-2015,Huang-arXiv-2022}. Owing to relatively simpler energy level structures compared to many proposed heavier HCI clock candidates,
the shortlisted candidates are easier to interrogate in the experiment and it is possible to carry out calculations of their spectroscopic
properties using the available state-of-the-art many-body methods. The clock transitions in these HCIs also show reasonably enhanced sensitivity
coefficients for testing local Lorentz symmetry invariance (LLI) and probing variation of $\alpha$. To demonstrate low systematic effects in the
clock frequency measurements, we have estimated major systematic effects due to the Zeeman, Stark, black-body radiation (BBR), and electric quadrupole
shifts by considering typical orders of field strengths.

\begin{figure}[t]
	\begin{center}
		\includegraphics[width=4.2cm]{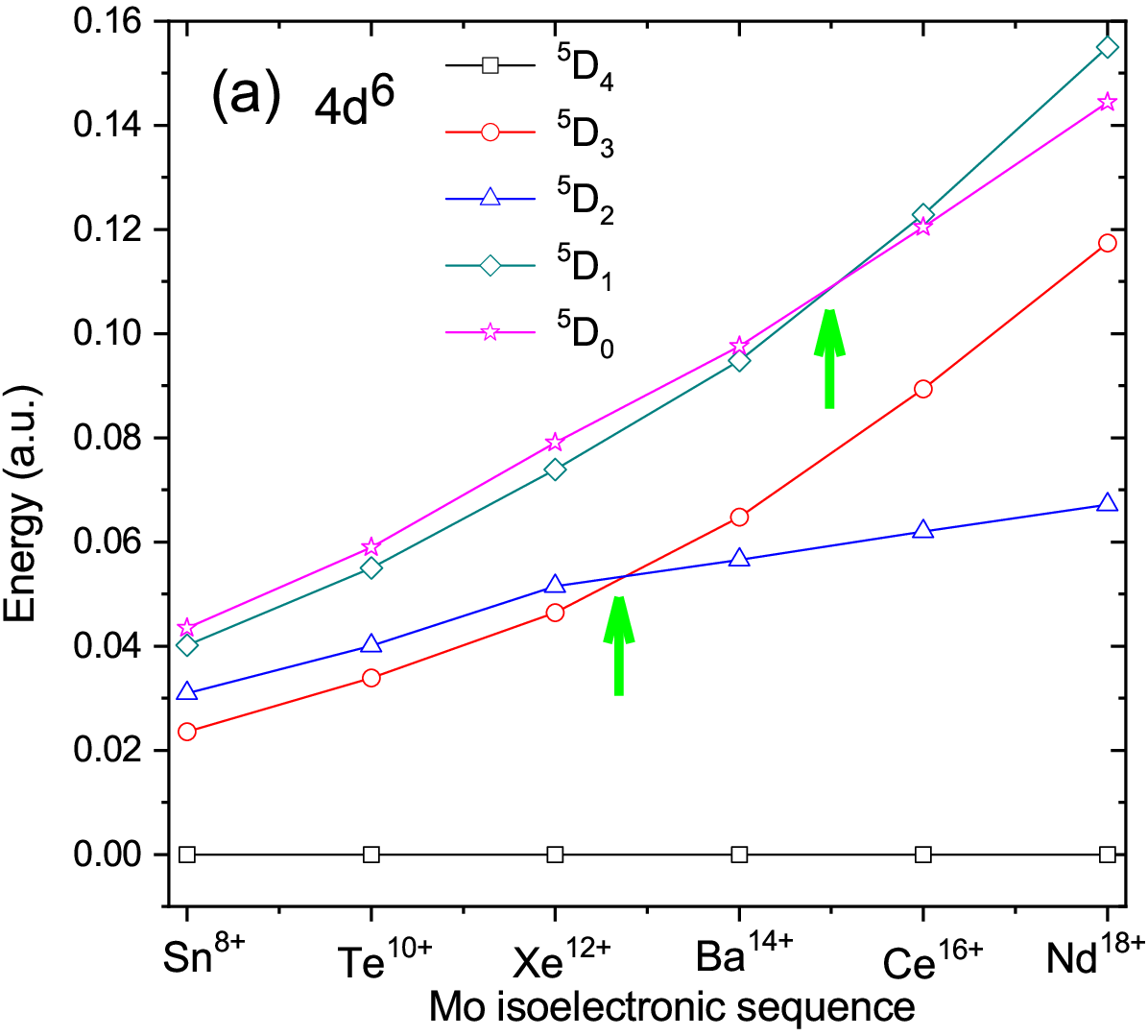}	
		\includegraphics[width=4.2cm]{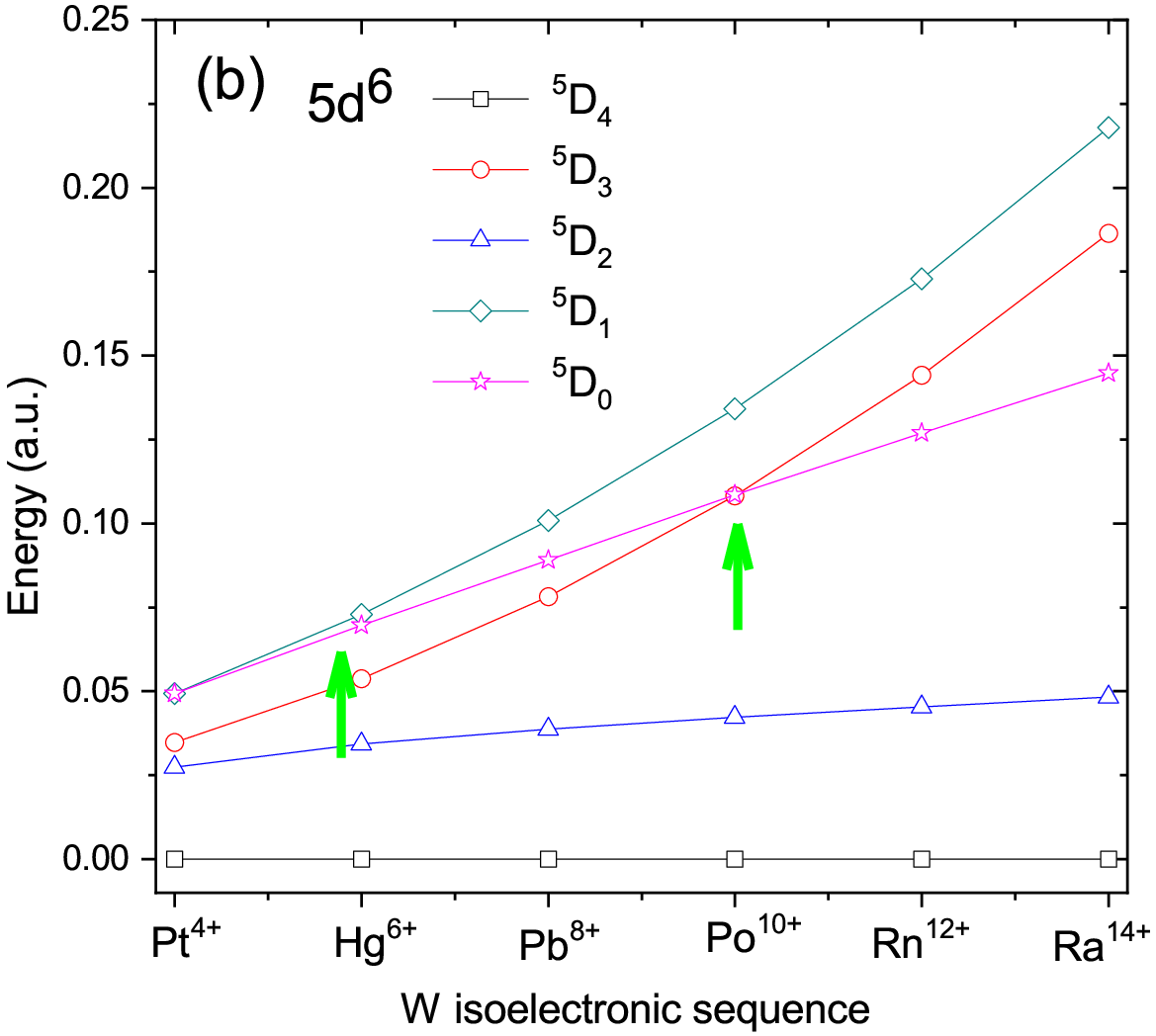}
		\includegraphics[width=4.2cm]{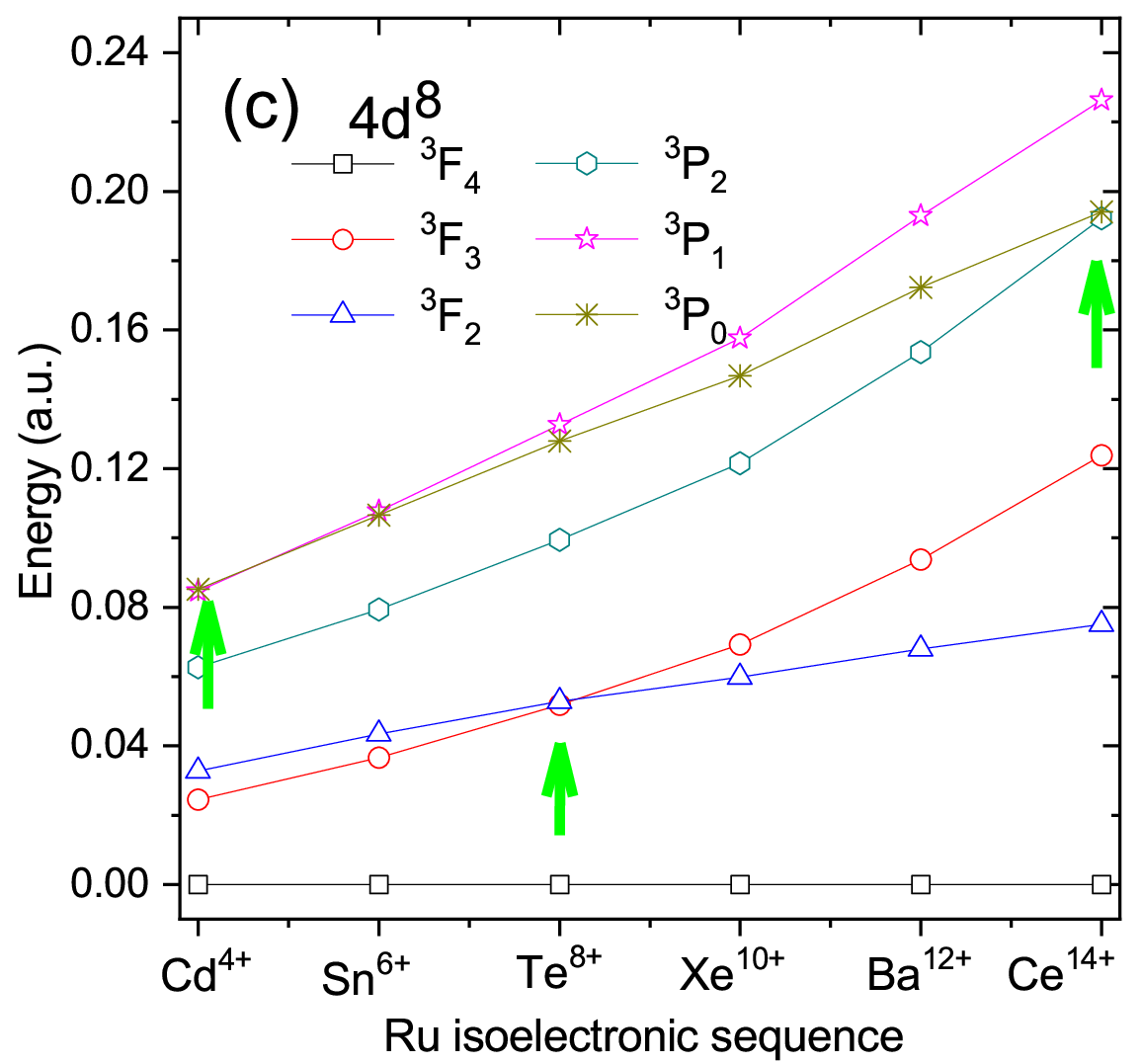}
		\includegraphics[width=4.2cm]{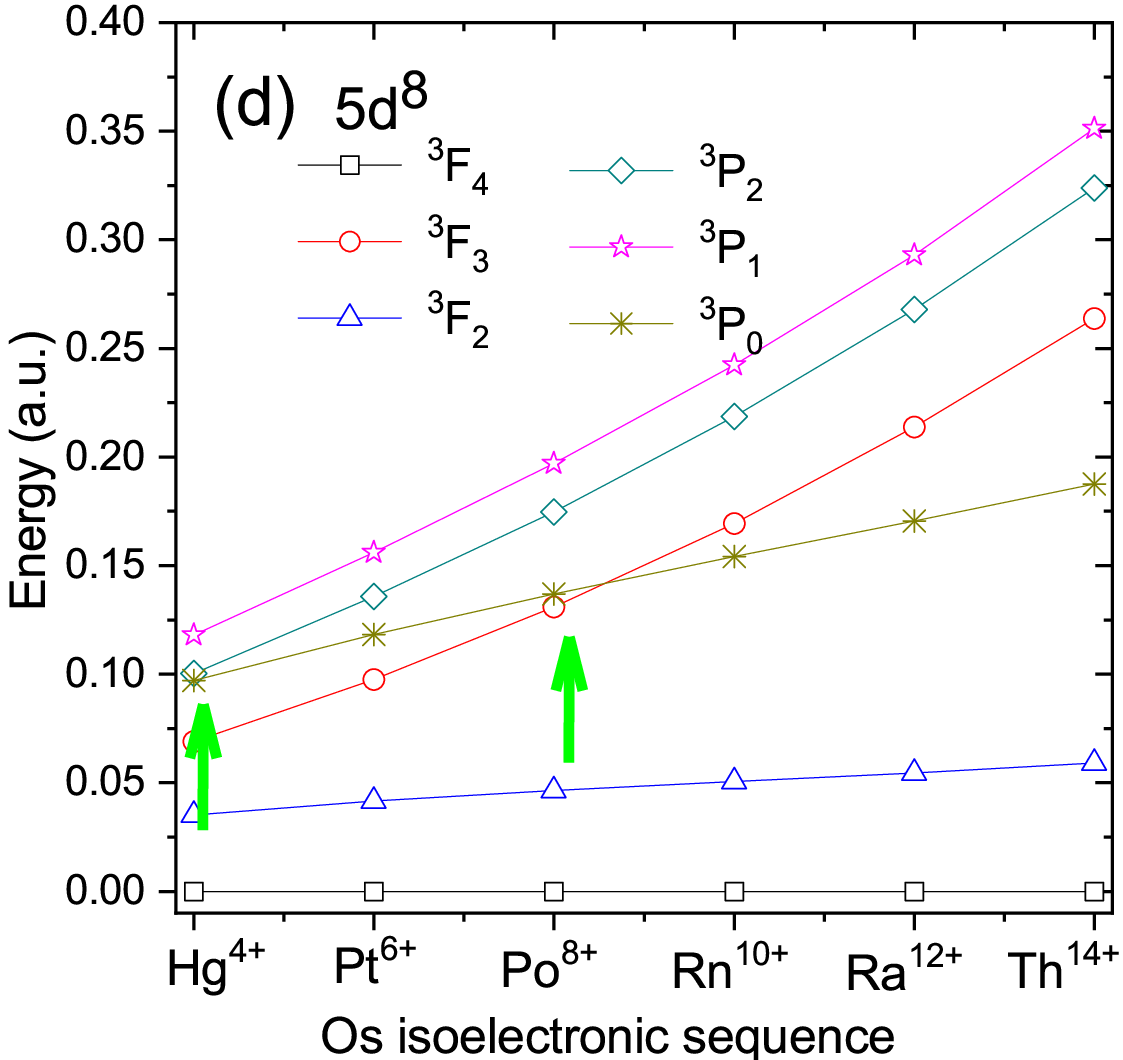}
	\end{center}
	\caption{Plots showing FS-ELCs (pointed out by arrows) in HCIs with the $(n=4,5)d^{6,8}$ configurations of the (a) Mo-like,
	(b) W-like, (c) Ru-like, and (d) Os-like isoelectronic sequences. Energies are given in atomic units (a.u.).
}
	\label{fig:EEcross}
\end{figure}

\begin{table}[t]
\caption{Trends of FS splitting energies (in cm$^{-1}$) in different isoelectronic sequences of the HCIs with the $n(=4,5)d^{6,8}$ configurations using
the CISDT method. ELCs occurring for the HCIs are highlighted by bold fonts. \label{tab:EEd6d8}}
	{\setlength{\tabcolsep}{4pt}
\begin{tabular}{l cccccc }\hline\hline
\multicolumn{7}{l}{ Mo-like~~~~$4d^6~^5D_4$ ground state  }													\\
	&	Sn$^{8+}$	&	Te$^{10+}$	&	Xe$^{12+}$	&	Ba$^{14+}$	&	Ce$^{16+}$	&	Nd$^{18+}$	\\
$^5D_3$	&	5161 	&	7441 	&	10190 	&	14223 	&	19619 	&	25759	\\
 $^5D_2$	&	6792 	&	8788 	&	11308 	&	\bf{12412}	&	\bf{13609}	&	\bf{14735}	\\
 $^5D_1$	&	8817 	&	12062 	&	16215 	&	20807 	&	26969 	&	34017	\\
 $^5D_0$	&	9543 	&	12937 	&	17357 	&	21412 	&	\bf{26444}	&	\bf{31706}	\\ \hline
\multicolumn{7}{l}{W-like~~~~$5d^6~^5D_4$ ground state}													\\
	&	Pt$^{4+}$	&	Hg$^{6+}$	&	Pb$^{8+}$	&	Po$^{10+}$	&	Rn$^{12+}$	&	Ra$^{14+}$	\\
 $^5D_3$	&	7412 	&	11792 	&	17160 	&	23749 	&	31649 	&	40925 	\\
 $^5D_2$	&	\bf{6637}	&	\bf{7553}	&	\bf{8488}	&	\bf{9261}	&	\bf{9944}	&	\bf{10617}	\\
 $^5D_1$	&	10756 	&	16002 	&	22124 	&	29425 	&	37949 	&	47811 	\\
 $^5D_0$	&	10893 	&	\bf{15271}	&	\bf{19546}	&	\bf{23814}	&	\bf{27867}	&	\bf{31777}	\\ \hline
\multicolumn{7}{l}{ Ru-like~~~~$4d^8~^3F_4$ ground state}													\\
	&	Cd$^{4+}$	&	Sn$^{6+}$	&	Te$^{8+}$	&	Xe$^{10+}$ 	&	Ba$^{12+}$	&	Ce$^{14+}$	\\
$^3F_3$	&	5378 	&	8034 	&	11378 	&	15114	&	20573 	&	27182 	\\
 $^3F_2$	&	7189 	&	9526 	&	11594 	&	\bf{13322}	&	\bf{14933}	&	\bf{16471}	\\
 $^3P_2$	&	13775 	&	17424 	&	21827 	&	26812	&	33724 	&	42210 	\\
$^3P_1$	&	18584 	&	23658 	&	29127 	&	34879	&	42360 	&	49661 	\\
$^3P_0$	&	18717 	&	\bf{23402}	&	\bf{28074}	&	\bf{32637}	&	\bf{37821}	&	\bf{42610}	\\ \hline
\multicolumn{7}{l}{Os-like~~~~$5d^8~^3F_4$ ground state}													\\
	&	Hg$^{4+}$	&	Pb$^{6+}$	&	Po$^{8+}$	&	Rn$^{10+}$	&	Ra$^{12+}$	&	Th$^{14+}$	\\
$^3F_3$	&	15142 	&	21409 	&	28721 	&	37189	&	46908	&	57897 	\\
 $^3F_2$	&	\bf{7732}	&	\bf{9104}	&	\bf{10170}	&	\bf{11096}	&	\bf{11960}	&	\bf{12953}	\\
 $^3P_2$	&	21996 	&	29778 	&	38349 	&	47965 	&	58781 	&	71044 	\\
 $^3P_1$	&	25876 	&	34245 	&	43202 	&	53149 	&	64286 	&	77057 	\\
 $^3P_0$	&	\bf{21296}	&	\bf{25957}	&	\bf{30041}	&	\bf{33816}	&	\bf{37407}	&	\bf{41165}	\\\hline\hline
	\end{tabular}}
\end{table}

\begin{table*}[t]
\caption{The estimated $\lambda$, $\tau$, $Q$, $\Delta R$, and $\langle J \|{\bf T}^{(2)} \| J \rangle$ values for the identified Clock-I and Clock-II
transitions through ELC in a large number of HCIs. The $\langle J \|{\bf T}^{(2)} \| J \rangle$ values for the ground and excited states of the only
Clock-I transition are given for representation. See text for definitions of notations. $a[b]$ denotes $a\times10^{b}$. \label{tab:ionclock}}
	{\setlength{\tabcolsep}{6pt}
\begin{tabular}{cccc cccc cccc ccc}\hline\hline
	&	\multicolumn{5}{c}{Clock-I }									&&	\multicolumn{5}{c}{Clock-II }									&&	 \multicolumn{2}{c}{$\langle J \|{\bf T}^{(2)} \| J \rangle$(a.u.) }			\\\cline{2-6}\cline{8-12} \cline{14-15}
	&	$\lambda$ (nm)	&	$\tau$ (s)	&	$Q$	&	$K_{\alpha}$	&	$\Delta R$	&&	$\lambda$ (nm)	&	$\tau$ (s)	&	$Q$	&	$K_{\alpha}$	&	$\Delta R$	&&	Ground	& Excited	\\\hline
\multicolumn{10}{l}{ Mo-like~~~~$4d^6$}																									\\
\bf{Nd$^{18+}$}	&	\bf{679}	&	\bf{860}	&	2.4[18]	&	\bf{0.1}	&	\bf{0.8}	&&	\bf{315}	&	\bf{138}	&	8.2[17]	&	\bf{1.5}	&	 \bf{2.1}	&&	\bf{147}	&	\bf{$-47.5$}	\\
\multicolumn{10}{l}{ W-like~~~~$5d^6$}																									\\
Hg$^{6+}$	&	1255 	&	1889 	&	2.8[18]	&	0.4	&	1.7 	&&	655 	&	505 	&	1.5[18]	&	1.7	&	3.3 	&&	51.2 	&	$-14.6$ 	\\
Pb$^{8+}$	&	1178 	&	1250 	&	2.0[18]	&	0.5	&	1.2 	&&	512 	&	107.0 	&	3.9[17]	&	1.3	&	2.5 	&&	62.2 	&	$-20.8$ 	\\
Po$^{10+}$	&	1080 	&	967 	&	1.7[18]	&	$-0.1$	&	1.1 	&&	420 	&	32.9 	&	1.5[17]	&	1.1	&	2.1 	&&	73.5 	&	$-27.9$ 	\\
\bf{Rn$^{12+}$}	&	\bf{1006}	&	\bf{810}	&	1.4[19]	&	$-$\bf{0.04}	&	\bf{0.9}	&&	\bf{359}	&	\bf{13.5}	&	7.1[16]	&	\bf{0.9}	&	 \bf{1.8}	&&	\bf{85.5}	&	$-$\bf{36.0}	\\
Ra$^{14+}$	&	942 	&	690 	&	1.3[18]	&	$-0.01$	&	0.9 	&&	315 	&	6.8 	&	4.1[16]	&	0.8	&	1.6 	&&	98.0 	&	$-44.7$ 	\\
\multicolumn{10}{l}{ Ru-like~~~~$4d^8$}																									\\
Te$^{8+}$	&	863 	&	772	&	1.7[18]	&	0.7	&	1.4 	&&	356 	&	14 	&	7.2[16]	&	0.5	&	1.3 	&&	$-41.7$ 	&	15.3 	\\
\bf{Xe$^{10+}$}	&	\bf{761}	&	\bf{480}	&	1.2[18]	&	\bf{0.4}	&	\bf{1.0}	&&	\bf{310}	&	\bf{9.3}	&	5.6[16]	&	\bf{0.5}	&	 \bf{1.3}	&&	\bf{$-52.1$}	&	\bf{31.6}	\\
Ba$^{12+}$	&	670 	&	285 	&	8.0[17]	&	0.2	&	0.9 	&&	264 	&	5.1 	&	3.6[16]	&	0.5	&	1.2 	&&	$-63.5$ 	&	46.8 	\\
\multicolumn{10}{l}{ Os-like~~~~$5d^8$}																									\\
Hg$^{4+}$	&	1293 	&	643 	&	9.4[18]	&	$-0.3$	&	1.8 	&&	470 	&	8.2 	&	3.3[16]	&	0.3	&	2.4 	&&	$-27.8$ 	&	23.4 	\\
\bf{Pb$^{6+}$}	&	\bf{1098}	&	\bf{393}	&	6.7[18]	&	$-$\bf{0.3}	&	\bf{1.5}	&&	\bf{385}	&	\bf{3.8}	&	1.9[16]	&	\bf{0.3}	&	 \bf{1.9}	&&	\bf{$-36.7$}	&	\bf{31.5}	\\
Po$^{8+}$	&	983 	&	311 	&	6.0[18]	&	$-0.1$	&	1.2 	&&	333 	&	2.3 	&	1.3[16]	&	0.3	&	1.5 	&&	$-46.7$ 	&	40.3 	\\
Rn$^{10+}$	&	901 	&	268 	&	5.6[19]	&	$-0.1$	&	1.1 	&&	296 	&	1.6 	&	1.0[16]	&	0.3	&	1.4 	&&	$-57.3$ 	&	49.2 	 \\\hline\hline
	\end{tabular}}
\end{table*}

\begin{table}[t]
\caption{A summary of properties relevant to systematic effects estimation for the clock states of a few representative HCIs evaluated
using the CISDT method. \label{tab:prop}}
{\setlength{\tabcolsep}{3pt}
\begin{tabular}{lcccc}\hline\hline
Items	&	Nd$^{18+}$	&	Rn$^{12+}$	&	Xe$^{10+}$	&	Pb$^{6+}$	\\ \hline
\multicolumn{5}{l}{\underline{Level configurations}} \\
Ground state	&	$4d^6~^5D_4$	&	$5d^6~^5D_4$	&	$4d^8~^3F_4$	&	$5d^8~^3F_4$	\\
Clock-I	state &	$4d^6~^5D_2$	&	$5d^6~^5D_2$	&	$4d^8~^3F_2$	&	$5d^8~^3F_2$	\\
Clock-II state	&	$4d^6~^5D_0$	&	$5d^6~^5D_0$	&	$4d^8~^3F_0$	&	$5d^8~^3F_0$	\\
\multicolumn{5}{l}{ \underline{Lande $g_J$-factors}} \\								
Ground state	&	1.4068	&	1.3387	&	1.2421 	&	1.2351 	\\
Clock-I	state &	1.406	&	1.3492	&	0.9994 	&	1.1171 	\\
\multicolumn{5}{l}{\underline{$\Theta$ values in a. u.}} 	\\
Ground state	&	0.1088 	&	0.2356 	&	-0.1190 	&	0.2828 	\\
Clock-I	&	-0.0326 	&	-0.1108 	&	0.0674 	&	0.2011 	\\
\multicolumn{5}{l}{\underline{$\alpha_{0}^{E1}$ values in a.u.}} \\
Ground state	&	0.2196 	&	0.8067 	&	0.5817 	&	2.4427 	\\
Clock-I	state &	0.2203 	&	0.8099 	&	0.5824 	&	2.4495 	\\
Clock-II state	&	0.2196 	&	0.8042 	&	0.5826 	&	2.4540 	\\
\multicolumn{5}{l}{\underline{$\delta \alpha_{0}^{E1}$ values in a.u. }} \\						
Clock-I transition	&	0.0007 	&	0.0032 	&	0.0007 	&	0.0069 	\\
Clock-II transition	&	0.0001 	&	-0.0025 	&	0.0009 	&	0.0113 	\\
\multicolumn{5}{l}{\underline{$\alpha_{2}^{E1}$ values in a.u.}}	\\					
Ground state	&	0.0047 	&	-0.0021 	&	0.0036 	&	0.0549 	\\
Clock-I	state &	-0.0015 	&	0.0031 	&	-0.0057 	&	-0.0291 \\
\multicolumn{5}{l}{\underline{$\alpha^{M1}$ values in a.u.}} \\									
Ground state	&	-5.5$\times10^4$	&	-3.8$\times10^4$	&	-5.7$\times10^{4}$	&	-5.5$\times10^{4}$	\\
Clock-I	state &	-2.9$\times10^5$	&	-1.1$\times10^5$	&	-3.0$\times10^{5}$	&	-6.4$\times10^{4}$	\\
Clock-II state	&	-4.8$\times10^6$	&	-4.5$\times10^5$	&	-2.1$\times10^{6}$	&	-5.3$\times10^{5}$	\\
\hline\hline
\end{tabular}}
\end{table}

To find out the energy level positioning of ions with $d^{6,8}$ configurations for laser trapping and interrogation purpose, we have employed three
independent first-principle many-body methods to calculate excitation energies (EE). We have adopted the following procedure to ensure about validity
of these calculations. First, we consider a representative HCI of a given isoelectronic sequence, for which experimental values are available, to
identify their ground and excited state configurations. Then, different many-body methods are employed to reproduce the EEs and cross-verify among them.
The initial ions under investigations are Cr-like Zn$^{6+}$, Fe-like Rb$^{11+}$, Mo-like Pd$^{4+}$, Ru-like Xe$^{10+}$, and W-like Pt$^{4+}$. Both the experimental values \cite{NIST,Raassen-PC-1986} and calculations from three independent methods of EEs of these HCIs
are given in Table \ref{tab:EEcomp}. We employ first the combined configuration interaction method and many-body perturbation theory (CI$+$MBPT) using
AMBiT code \cite{Kahl-CPC-2019} in which correlations among the valence electrons from the $n(=3,4,5)d^{6,8}$ orbitals are treated explicitly at the
singles and doubles excitation approximation in the CI method and correlations among the valence electrons and the atomic core are accounted for using
the MBPT method. We find good agreement between the CI$+$MBPT results and the experimental data for the systems with $n=3$, but large
differences are seen for the ions with $n=4,5$. Next, we adopt the singles and doubles approximated Fock-space coupled-cluster theory (CCSD method)
\cite{Visscher-JCP-2001} implemented in the DIRAC program \cite{Dirac} to obtain the FS among the states with $nd^8$ configurations by detaching two
electrons from the closed-core $nd^{10}$. The CCSD results show similar accuracy with the CI$+$MBPT results. Then, we applied the multi-reference configuration interaction method with the singles and doubles excitations (CISD method) followed by singles, doubles, and triples excitations (CISDT method) in the general active space scheme \cite{Fleig-JCP-2001, Fleig-JCP-2003,
Fleig-JCP-2006} that are implemented in the DIRAC program \cite{Knetch-JCP-2008, Dirac}. In both the CISD and CISDT methods, electrons only from the
valence $nd$ shells and from the $(n-1)s,p$ shells are allowed in the excitation processes. In most of heavier HCIs, the CISDT results are found to be
better in agreement with the experimental values than the CI$+$MBPT and CCSD results. Thus, results from the CISDT method, that incorporates more physical
effects through the triple excitations than the CISD method, are considered as the final calculated values. Nonetheless, comparative computations using
all the three considered many-body methods helped us to test reliability of the calculations.

We analyze the FS-ELCs of the aforementioned isoelectronic sequence HCIs with $nd^6$ and $nd^8$ configurations for $n=3,4$ and 5 by increasing ionization number
$Z_{\rm ion}$. We focus mainly on the heavier HCIs with moderate values of $Z_{\rm ion}$ lying in the range 4-20 that can be produced using table-top
electron ion beam facility. Furthermore, we concentrate excited states within the range of 300-1000nm that can be easily accessible by the available
lasers. The radioactive HCIs with $n(=6)d^{6,8}$ configurations having half-lifetimes shorter than seconds are not investigated because they will not be
suitable to consider in the laboratory for making clocks. The EEs between the FS splitting in the $n(=4,5)d^{6,8}$ isoelectronic sequences are shown in
Fig. \ref{fig:EEcross}. The $n(=3)d^{6,8}$ isoelectronic sequences are not shown since no ELCs are found for them. Quantitative data for EEs are tabulated
in Table \ref{tab:EEd6d8} from the CISDT method. Since the Mo-like and W-like isoelectronic sequences have $n(=4,5)d^{6}$ configurations,
they will have low-lying FS levels as $^5D_{J}$ for $J=0,1,2,3,4$. It can be seen from  Fig. \ref{fig:EEcross}(a) that for smaller $Z_{\rm ion}$ values
in the Mo-like isoelectronic sequences, such as from Sn$^{8+}$ to Xe$^{12+}$, the FS level ordering follows as $J=4,3,2,1,0$. For increasing
$Z_{\rm ion}$ value, the $^5D_3$ and $^5D_2$ levels swap the ordering around Ba$^{12+}$ and the $^5D_{0}$ level moves downwards to be below
the $^5D_{1}$ level around Ce$^{16+}$. Thus, it shows ELCs between the $J=2$ and $J=3$ levels then between the $J=0$ and $J=1$ levels as indicated by
arrows in Fig. \ref{fig:EEcross}(a) and shown in bold fonts in Table \ref{tab:EEd6d8}. The W-like isoelectronic sequences also show similar ELCs, as
shown in Fig. \ref{fig:EEcross}(b) and in the above table. The Ru-like and the Os-like isoelectronic sequences are found to have FS splitting as
$^3F_2$, $^3F_1$, $^3F_0$, $^3P_2$, $^3P_1$, and $^3P_0$ for lower $Z_{\rm ion}$ values, but these ordering changes from $^3F_3$ to $^3F_2$,
from $^3P_1$ to $^3P_0$, then from $^3P_0$ to $^3F_3$ as shown by arrows in Figs. \ref{fig:EEcross}(c) and (d) for which data are quoted in Table \ref{tab:EEd6d8}.

After understanding ELCs in different isoelectronic systems investigated here, we now intend to use this knowledge in identifying potential candidates
for atomic clocks. We find at least two clock transitions in each type of isoelectronic HCIs showing ELCs, which are named as Clock-I and Clock-II
for further discussions. In HCIs with $nd^6$ configurations, the $^5D_4-{^5D_2}$ and $^5D_4-{^5D_0}$ ELC transitions basically correspond to the Clock-I
and Clock-II transitions respectively. Similarly, the $^3F_4-{^3F_2}$ and $^3F_4-{^3P_0}$ ELC transitions in the ions with the $nd^8$ configurations
can be the Clock-I and Clock-II transitions respectively. Since these forbidden transitions are mainly guided by the electric quadrupole (E2) decay
channel, they are expected to be quite narrow leading to very long lifetime for the excited states. We have verified this by estimating the decay rates
and lifetimes of the excited states. In Table \ref{tab:ionclock}, we list a number of HCIs that have Clock-I and Clock-II transitions in the wavelength
($\lambda$) around 300-1000 nm. We provide the $\lambda$ values and $Q$ values of the clock transitions and lifetimes ($\tau$) of the excited states
associated with both the clock transitions. As can be noticed from this table, the $\tau$ and $Q$ values in almost all the listed HCIs are sufficiently
large. This essentially advocates that they all are apt for making ultra-stable optical clocks, but we have highlighted some of the HCIs in bold fonts
in the above table which, we feel, have overall advantages for clocks in view of laboratory consideration and probing fundamental physics.

The $\alpha-$variation sensitivity coefficients are defined as
\begin{eqnarray}
K_{\alpha}=2(q_2-q_1)/h\nu ,
\end{eqnarray}
where $q_{1/2}$ are $\alpha-$variation sensitivity coefficients of the ground and excited states of the clock transition respectively, $h$ is the Plank
constant, and $\nu$ is the clock frequency. The $K_{\alpha}$ values of the proposed clock transitions presented in Table \ref{tab:ionclock} show that
they are comparable or larger than the $^{171}$Yb clock transitions \cite{Tang-PRA-2023}. Similarly, the LLI
interaction Hamiltonian is given by \cite{Hohensee-PRL-2013}
\begin{eqnarray}
 H^{LLI} = -C_0^{(0)} \frac{\textbf{p}^2}{2 m_e} - C_0^{(2)} \frac{\textbf{T}^{(2)}}{6 m_e} ,
\end{eqnarray}
where $m_e$ is the mass of an electron, $C_0^{(0/2)}$ are related to LLI violating coefficients, $\textbf{p}$ is the momentum operator and $\textbf{T}^{(2)}$ is a second-rank tensor. Therefore, enhancement of LLI in an atomic transition depends on the kinetic energy (K.E.) and sensitivity of the $\textbf{T}^{(2)}$ operator in the states involved. Enhancement due to K.E. of electrons are defined as $\Delta R=-(\langle J \|{\bf p}^2 \| J \rangle_2-\langle J \|{\bf p}^2 \| J
\rangle_1)/(2h\nu)$, while $\textbf{T}^{(2)}$ are estimated by the expectation values. Clock-I transitions highlighted in Table \ref{tab:ionclock} show
very large differential values for $\Delta R$ and expectation values of the $\textbf{T}^{(2)}$ operator. Though these values are slightly smaller than the
octupole transition of Yb$^+$ \cite{Dzuba-NP-2016}, but they are much larger than the clock transitions of Ca$^+$ \cite{Sahoo-PRA-2019} and Sr$^+$ \cite{Shaniv-PRL-2018}. It indicates that the proposed clock transitions are not only suitable to be considered for ultra-precise clocks they are also very sensitive to probe LLI violation.

Next, we discuss the major systematic effects in the clock transitions of a few representative Nd$^{18+}$, Rn$^{12+}$, Xe$^{10+}$, and Pb$^{6+}$ HCIs as
these effects are of almost similar orders in other considered HCIs. Among these, it would be easier to consider the Xe$^{10+}$ ion in the laboratory as
proof-of-principle for measuring its energy levels are already demonstrated. To estimate typical orders shifts due to electric fields and BBR shifts in the above
HCIs, we have determined scalar ($\alpha_0^{E1}$) and tensor ($\alpha_2^{E1}$) components of $\alpha^{E1}$, and E2 moments ($\Theta$) of the ions using
the CISDT method in the finite-field (FF) approach. As seen in Table \ref{tab:prop}, the $\delta \alpha_0^{E1}$ values are around $10^{-3} - 10^{-4}$. It means that the fractional differential Stark shifts, $\delta E_{Stark}$=$-\delta \alpha_0^{E1} {\cal E}^2/2$, of the clock transitions in the above HCIs, for a typical electric field strength ${\cal E}$=$10 \ V/m$, can be
negligibly small. Contributions from the $\alpha_2^{E1}$ values can be minimized by choosing suitable quantization axis while carrying out the
measurements. Also for such small $\delta \alpha_0^{E1}$ values, the BBR shifts can be suppressed far below $10^{-19}$ in the cryogenic
environments. The excited state of Clock-II transition in Rn$^{12+}$ has negative $\delta \alpha_0^{E1}$ value, which can help in producing ``magic"
trap drive frequency, $\Omega$, as mentioned in Ref. \cite{Dube-PRA-2013} and estimated to be $\Omega\approx2\pi \times 520$ MHz. The electric quadrupole
shift, given by $\Delta E_{Quad}=-\Theta \mathcal E_{zz}/2$ with $\mathcal E_{zz}$ is the gradient of the applied electric field in the $z$-direction,
found to be around $10^{-14} - 10^{-15}$ in the above HCIs. It is possible to nullify uncertainties due to these shifts by performing frequency
measurements in all azimuthal $M$-components and considering the averaged frequency. We have also determined the Lande $g_J$-factors and the scalar M1
polarizabilities ($\alpha_0^{M1}$) in the expectation value approach using the CISDT method. The $g_J$ values will be useful for calibrating
the magnetic field strengths ($B$) during the measurements. The first-order Zeeman shifts can be eliminated by choosing the $M_J=0\rightarrow M_J=0$ transitions
or adopting the averaging frequency measurement technique. The second-order Zeeman shift can be estimated by $\delta E^{(2)}_{Zeem}=
-\frac{1}{2} \delta\alpha^{M1}B^2$, where $\delta\alpha^{M1}$ is the differential $\alpha^{M1}$ values for the clock transitions. Assuming a typical
magnetic field strength of $B=5\times10^{-8}$T, the fractional frequency shifts, $\delta E^{(2)}_{Zeem}/\nu$, are found to be below $10^{-19}$. Thus,
the aforementioned discussions suggest that the fractional uncertainties to the proposed Clock-I and Clock-II transitions in the investigated HCIs
can be $10^{-19}$ level. Nonetheless, some of these systematic effects can be further reduced by selecting transitions among hyperfine levels of the
clock states appropriately.

In Summary, we have identified a large number of heavier highly charged ions by analyzing energy level-crossings of the fine-structure splitting of
the ground $d^6$ and $d^8$ open-shell configurations that are advantageous for making ultra-stable and high-precision optical atomic clocks. They
offer at least two set of clock transitions with quality factors about $10^{16-18}$ and fractional uncertainties due to major systematics effects
are below $10^{-19}$ level compared to the currently undertaken Ar$^{13+}$ clock transition. These clock transitions also show very sensitive
for probing fundamental phenomena like possible temporal variation of the fine structure constant and local Lorentz symmetry invariance.

YY thanks K. Yao for sharing the work of the $3d$ ion EBIT spectroscopy. This work is supported by The National Key Research and Development Program
of China (2021YFA1402104), and Project supported by the Space Application System of China Manned Space Program. BKS would like to acknowledge use of
ParamVikram-1000 HPC of Physical Research Laboratory (PRL), Ahmedabad.


\begin{thebibliography}{}

\bibitem{King-nature-2022}
S. A. King, L. J. Spie$\beta$, P. Micke, A. Wilzewski, T. Leopold, E. Benkler, R. Lange, N. Huntemann, A. Surzhykov, V. A. Yerokhin, J. R.
Crespo L${\acute{o}}$pez-Urrutia, P. O. Schmidt, Nature {\bf 611}, 43 (2022).

\bibitem{Yudin-PRA-2014}
V. I. Yudin, A. V. Taichenachev, and A. Derevianko, Phys. Rev. Lett. {\bf 113}, 233003 (2014).

\bibitem{Yu-PRA-2019}
Y. M. Yu and B. K. Sahoo, Phys. Rev. A {\bf 99}, 022513 (2019).

\bibitem{Safronova2018}
M. S. Safronova, D. Budker, D. DeMille, Derek F. Jackson Kimball, A. Derevianko, and C. W. Clark, Rev. Mod. Phys. {\bf 90}, 025008 (2018).

\bibitem{Kozlov2018}
M. G. Kozlov, M. S. Safronova, J. R. Crespo Lopez-Urrutia, and P. O. Schmidt, Rev. Mod. Phys. {\bf 90}, 045005 (2018).

\bibitem{Yu2023}
Y. M. Yu, B. K. Sahoo, and B. B. Suo, Front. Phys. {\bf 11}, 1104848 (2023).

\bibitem{Robiscoe-PR-1965}
R. T. RoszscoEf, Phys. Rev {\bf 138}, A22, (1965).

\bibitem{Levine-RPL-1969}
J. S. Levine, P. A. Bonczyk, and A. Javan, Phys. Rev. Lett. {\bf 22}, 267 (1969).

\bibitem{Berengut-RPL-2010}
J. C. Berengut, V. A. Dzuba, and V.V. Flambaum, Phys. Rev. Lett. {\bf 105}, 120801 (2010).

\bibitem{Berengut-PRA-2012}
J. C. Berengut, V. A. Dzuba, V. V. Flambaum, and A. Ong, Phys. Rev. A {\bf 86}, 022517 (2012).

\bibitem{Berengut-PRL-2011}
J. C. Berengut, V. A. Dzuba, V.V. Flambaum, and A. Ong, Phys. Rev. Lett. {\bf 106}, 210802 (2011).

\bibitem{Bekker-NC-2019}
H. Bekker, A. Borschevsky, Z. Harman, C. H. Keitel, T. Pfeifer, P. O. Schmidt, J. R. Crespoez-Urrutia, J. C. Berengut, Nat. Commun. {\bf 10}, 5651 (2019).

\bibitem{Berengut-PRL-2012}
J. C. Berengut, V. A. Dzuba, V.V. Flambaum, and A. Ong, Phys. Rev. Lett. {\bf 109}, 070802 (2012).

\bibitem{Porsev-PRA-2020}
S. G. Porsev, U. I. Safronova,M. S. Safronova, P. O. Schmidt, A. I. Bondarev, M. G. Kozlov, I. I. Tupitsyn, and C. Cheung, Phys. Rev. A {\bf 102}, 012802 (2020).

\bibitem{Yu-atom-2022}
Y. M. Yu, P. Duo, S. L. Chen, B. Arora, H. Guan, K. Gao, and J. Chen, Atoms {\bf 10}, 123 (2022).

\bibitem{Lange-PRL-2021}
R. Lange, N. Huntemann, J. M. Rahm, C. Sanner,H. Shao, B. Lipphardt , Chr. Tamm, S. Weyers, and E. Peik, Phys. Rev. Lett. {\bf 126}, 011102 (2021).

\bibitem{Arnold-PRA-2015}
K. Arnold, E. Hajiyev, E. Paez, C. H. Lee, M. D. Barrett, and J. Bollinger,Phys. Rev. A {\bf 92}, 032108 (2015).

\bibitem{Huang-arXiv-2022}
Y. Huang, H. Guan, C. Li, H. Zhang, B. Zhang, M. Wang, L. Tang, T. Shi, and K. Gao, unpublished (arXiv 2022.07828 (2015)).

\bibitem{NIST}
A. Kramida, Y. Ralchenko, J. Reader, and NIST ASD Team https://physics.nist.gov/asd.

\bibitem{Raassen-PC-1986}
A.J.J. Raassen and Th. A. M. van Kleef, Physica C {\bf 142}, 359 (1986).

\bibitem{Kahl-CPC-2019}
E. V. Kahl and J. C. Berengut, Comp. Phys. Comm. {\bf 238}, 232 (2019).

\bibitem{Visscher-JCP-2001}
L. Visscher, E. Eliav, and U. Kaldor, J. Chem. Phys. {\bf 115}, 9720 (2001).

\bibitem{Dirac}
DIRAC, a relativistic ab initio electronic structure program, Release DIRAC22 (2022),
written by H. J. Aa. Jensen, et al. (available at http://dx.doi.org/10.5281/zenodo.6010450, see also http://www.diracprogram.org).

\bibitem{Fleig-JCP-2001}
T. Fleig, J. Olsen and C. M. Marian, J. Chem. Phys. {\bf 114}, 4775 (2001).
		
\bibitem{Fleig-JCP-2003}
T. Fleig and J. Olsen and L. Visscher, J. Chem. Phys. {\bf 119}, 2963 (2003).
		
\bibitem{Fleig-JCP-2006}
T. Fleig and H. J. A. Jensen and J. Olsen and L. Visscher, J. Chem. Phys. {\bf 124}, 104106 (2006).
		
\bibitem{Knetch-JCP-2008}
S. Knecht, H. J. A. Jensen, and T. Fleig, J. Chem. Phys. {\bf 128}, 014108 (2008).

\bibitem{Tang-PRA-2023}
Z. M. Tang, Y. M. Yu, B. K. Sahoo, C. Z. Dong, Y. Yang, and Y. Zou, Phys. Rev. A {\bf 107}, 053111 (2023).

\bibitem{Hohensee-PRL-2013}
M. A. Hohensee, N. Leefer, D. Budker, C. Harabati, V. A. Dzuba, and V. V. Flambaum, Phys. Rev. Lett. {\bf 111}, 050401 (2013).

\bibitem{Dzuba-NP-2016}
V. A. Dzuba, V. V. Flambaum, M. S. Safronova, et al., Nat. Phys. {\bf 12}, 465 (2016).

\bibitem{Sahoo-PRA-2019}
B. K. Sahoo, Phys. Rev. A {\bf 99}, 050501(R) (2019).

\bibitem{Shaniv-PRL-2018}
R. Shaniv, R. Ozeri, M. S. Safronova, S. G. Porsev, V. A. Dzuba, V. V. Flambaum, and H. H\"{a}ffner Phys. Rev. Lett. {\bf 120}, 103202 (2018)

\bibitem{Dube-PRA-2013}
P. Dub\'{e}, A. A. Madej, Z. Zhou, and J. E. Bernard, Phys. Rev. A {\bf 87}, 023806 (2013).

\end{thebibliography}
\end{document}